\documentclass[12pt]{article}
\usepackage[backend=biber, style=nature]{biblatex}
\usepackage{color}
\usepackage{times}
\usepackage{graphicx} 
\usepackage{authblk}
\usepackage{lineno}
\usepackage{subcaption}
\usepackage{xcolor}
\usepackage{geometry}
\usepackage{amsmath}
\usepackage{amssymb}
\usepackage{tikz}
\usepackage[nottoc,numbib]{tocbibind}
\usepackage{titletoc}
\usepackage{mathtools}
\usepackage{braket}
\usepackage{subcaption}
\usepackage{subfloat}
\usepackage{cleveref}
\usepackage{upgreek}
\usepackage[separate-uncertainty = true,multi-part-units=single]{siunitx}
\makeatletter
\let\ftype@table\ftype@figure
\makeatother
\renewcommand{\thefigure}{\arabic{figure}}
\renewcommand{\figurename}{Fig.}
\renewcommand{\thetable}{\arabic{table}}
\renewcommand{\tablename}{Table}
\title{Tuneable entangled photon pair generation in a liquid crystal}

\author
{Vitaliy Sultanov,$^{1,2,\dagger}$ Aljaž Kavčič,$^{3,4,\dagger}$ Emmanuil Kokkinakis,$^{5}$ Nerea Sebastián,$^{3}$ Maria V. Chekhova,$^{1,2\ast}$ Matjaž Humar$^{3,4,6}$\\
\normalsize{$^{1}$Friedrich-Alexander Universität Erlangen-Nürnberg, 91058 Erlangen, Germany}\\
\normalsize{$^{2}$Max-Planck Institute for the Science of Light, Staudtstrasse 2 91058 Erlangen, Germany}\\
\normalsize{$^{3}$Jožef Stefan Institute, 1000 Ljubljana, Slovenia}\\
\normalsize{$^{4}$Faculty of Mathematics and Physics, University of Ljubljana, 1000 Ljubljana, Slovenia}\\
\normalsize{$^{5}$University of Crete, 71003 Heraklion, Greece}\\
\normalsize{$^{6}$CENN Nanocenter, 1000 Ljubljana, Slovenia}\\
\normalsize{$^{\dagger}$ These authors contributed equally}\\
\normalsize{$^\ast$To whom correspondence should be addressed; E-mail:  maria.chekhova@mpl.mpg.de.}
}

\date{}

\begin{document}

\baselineskip24pt

\maketitle

\begin{abstract}
   Liquid crystals, with their ability of self-assembly, strong response to the electric field, and integrability into complex systems, are key materials in light beam manipulation~\cite{khoo2022liquid}. Recently discovered ferroelectric nematic liquid crystals~\cite{nishikawa2017fluid,jmandle2017} also possess a considerable second-order optical nonlinearity, making them a perspective material for nonlinear optics~\cite{Folcia2022,Sebastian2021}. Their use as sources of quantum light could drastically extend the boundaries of photonic quantum technologies~\cite{Obrien}. However, spontaneous parametric down-conversion, the basic source of entangled photons~\cite{Kwiat1995}, heralded single photons~\cite{Hong1986}, and squeezed light~\cite{Vahlbruch2016}, has been so far not observed in liquid crystals - or in any liquids or organic materials. Here, we implement spontaneous parametric down-conversion in a ferroelectric nematic liquid crystal and demonstrate electric-field tunable broadband generation of entangled photons, with the efficiency comparable to that of the best nonlinear crystals. The emission rate and polarization state of photon pairs is dramatically varied by applying a few volts or twisting the molecular orientation along the sample. A liquid crystal source enables a new type of quasi-phasematching~\cite{zhao2022nontrivial}, which is based on molecular twist structure and is therefore reconfigurable for the desired spectral and polarization properties of photon pairs. Such sources promise to outperform standard nonlinear optical materials in terms of functionality, brightness and the tunability of the generated quantum state. The concepts developed here can be extended to complex topological structures, macroscopic devices, and multi-pixel tunable quantum light sources. 
\end{abstract}

\newpage

Liquid crystals (LCs) uniquely combine long-range molecular order and fluidity, which results in the self-assembly of various complex three-dimensional topological structures, birefringence, and large response to external stimuli~\cite{Gennes_Prost}. For this reason, LCs are employed in several active optical devices, notably liquid-crystal displays, tunable filters, spatial light modulators, and many others~\cite{Wu_Yang,khoo2022liquid}. Recently, ferroelectric nematic liquid crystals (FNLCs) have been discovered~\cite{nishikawa2017fluid,jmandle2017,chen2020first,sebastian2020ferroelectric}, which have polar ordering, leading to a large dielectric constant, a strong response to an electric field, and a very high optical nonlinear response. Among other possible uses, FNLCs have strong potential for applications in tunable nonlinear devices~\cite{Sebastian2021, Folcia2022}. Efficient second harmonic generation has been demonstrated~\cite{Folcia2022}, but the use of LCs as sources of quantum states of light has remained unexplored till now. 

Most quantum light sources rely on spontaneous four-wave mixing (SFWM) or spontaneous parametric down-conversion (SPDC). In SPDC, a single photon of a laser source (pump) is converted in a second-order nonlinear material into two daughter photons, which can be entangled in various degrees of freedom. Although SPDC was discovered half a century ago~\cite{Magde67,Harris67,Akhmanov68}, and has become the workhorse of quantum optics, sources of photon pairs and heralded photons based on it have barely evolved since then. The necessity to satisfy energy and momentum conservation laws (also known as the phase-matching condition) implies a careful source design and limits the set of two-photon states that can be generated. Existing solutions such as pump beam modulation~\cite{URen2019}, periodically ~\cite{Boyd} or aperiodically~\cite{Torres2012} poled nonlinear crystals or waveguides~\cite{Tanzilli2002} or holographic modulation~\cite{Arie2023} extend those limits but lack tunability and are designed for the generation of a particular quantum state. Integrated SPDC sources~\cite{Wang2020,Wang2021} are very promising for quantum technologies but still have the same restrictions.  The emerging field of quantum optical metasurfaces~\cite{Santigo-Cruz2022} pushes the boundaries of quantum state engineering but at the cost of significantly reduced generation efficiency.

\begin{figure}[hbt!]
    \centering
    \includegraphics[width=1\linewidth]{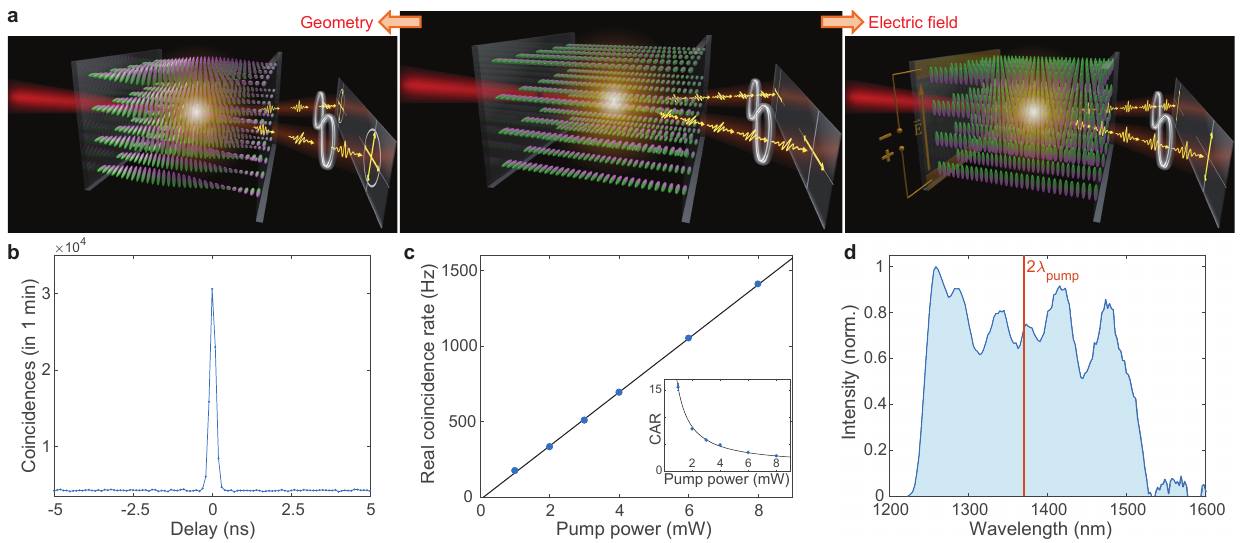}
    \caption{\textbf{Generation of entangled photons in a FNLC cell.} {\bf{a)}} The main concept of the work: both the flux and 
     the polarization state of the photon pairs 
     can be altered by reconfiguring the molecular orientation, achieved by either engineering the sample geometry or applying an electric field. {\bf{b)}} Typical peak of photon pair detection (coincidences) at the zero delay between the counts of two detectors. The $8\mu$m sample has no twist and SPDC is pumped with a \SI{6}{mW} CW laser at \SI{685}{nm}. {\bf{c)}} The rate of two-photon detection increases linearly with the power of the pump laser. Inset: the contrast of the peak (coincidence-to-accidentals ratio, CAR) versus the pump power. The inverse dependence indicates two-photon emission. {\bf{d)}} The spectrum of the generated photon pairs is broadband (limited by the cut-off wavelengths of the used filters) and relatively flat.}
    \label{overview}
\end{figure}

Meanwhile, FNLCs possess great potential for quantum optics as a key ingredient for the new generation of quantum light sources. The change of the FNLC's structure via an applied electric field offers a fine spatial tuning of the optical properties in real time. Here, we demonstrate an electrically tunable source of entangled photons based on SPDC in liquid crystals. This is the first observation of SPDC in liquid or any organic material; moreover, the rate is fairly high. We show that the two-photon polarization state can be altered via either a molecular orientation twist along the sample or an applied electric field  (Fig.~\ref{overview}a). We believe that this work lays the foundations for a new era of tunable quantum light sources.

\section*{Photon pair generation in liquid crystals}
The nonlinear properties of conventional solid crystals or waveguides used in nonlinear and quantum optics are fixed and defined by the orientation and structure. In contrast, the nonlinearity of FNLCs follows the orientation of the molecules, which can be very complex and can be manipulated by external stimuli. To investigate photon pair generation in a liquid crystal, we prepared several samples with different molecular orientations along the sample, possessing $0$-, $\pi/2$-, or $\pi$-twist of the molecules (Extended Data Fig.~1). The samples had high values of refractive index and birefringence (Extended Data Fig.~2, Section 1 of Supplementary Information) and a thickness of \SIrange{7}{8}{\micro m}, which was well below the nonlinear coherence length ($\sim\SI{19}{\micro m}$), and therefore generated photon pairs coherently~\cite{Okoth2019}. The sample with no twist had two electrodes with a gap of \SI{500}{\micro m} between them to create a uniform electric field, which reoriented the molecules (Fig.~\ref{overview}a). Therefore, we separately investigated how a predefined molecular twist and an electric-field-induced change of the molecular orientation affected the state of the generated photon pairs. 

Among the three degrees of freedom of two-photon light, such as position-momentum, time-frequency, and polarization, we focus here on polarization because its dependence on the molecules' orientation is most dramatic. With the axis $z$ defined along the molecular dipole moment, only one component of FNLC  second-order nonlinear tensor is significant: $d_{33}\approx\SI{20}{pm/V}$, only 40\% lower than for lithium niobate (Extended Data Fig.~3, Section 2 of Supplementary Information). Therefore, photon pairs are generated with the initial polarization along this axis, whose orientation is effectively equivalent to the orientation of the optic axis of a uniaxial crystal. The interference between all differential polarization states generated along the sample gives the resulting two-photon polarization state.

In each sample, SPDC is pumped with \SI{685}{nm} continuous-wave laser radiation up to a power of \SI{10}{mW} focused into a \SI{5}{\micro m} spot. At this pump power, we did not observe any other nonlinear effects or permanent damage to the sample (Section 3 of Supplementary Information). We detect photon pairs with a Hanbury Brown and Twiss interferometer looking at the correlations between the detection times of two photons (Extended Data Fig.~5, see Methods). Photons of the same pair arrive at the detectors simultaneously, creating a peak in the distribution of the time delay between two detection events (Fig.~\ref{overview}b). In contrast, uncorrelated photons from different pairs or generated via a different process (e.g., photoluminescence) lead to accidental coincidences, equally distributed over the delay times. While the number of photon pairs generated via SPDC depends linearly on the pump power (Fig.~\ref{overview}c), the ratio between the height of the peak and the background level of accidental coincidences (coincidence-to-accidental ratio, CAR) is inversely proportional to the pump power (the inset of Fig.~\ref{overview}c). The data shown in Fig.~\ref{overview} b,c clearly prove photon pair generation from a liquid crystal, with a fairly high coincidence rate. A narrow peak under CW pumping indicates time-frequency entanglement~\cite{Okoth2019}, but we do not quantify it here. 

Due to the microscale source thickness~\cite{Sultanov2022} and the resulting relaxed phase-matching condition, the spectrum of photon pairs from an FNLC layer should be broadband. We demonstrate it by measuring the two-photon spectrum (Fig.~\ref{overview}d) via two-photon fiber spectroscopy~\cite{Valencia2002}, accounting for the spectral detection efficiency and losses (Extended Data Fig.~6, see Methods). The generated two-photon spectrum is almost flat, up to a modulation caused by the etalon effect inside the source~\cite{Kitaeva2004}, and limited by the frequency filtering. Without filtering, the spectrum of photon pairs is expected to be even broader, suggesting applications like ultrafast time resolution, high-dimensional time/frequency quantum coding, or hyperentanglement. 

\section*{Photon pair polarization state switching with electric field}
\begin{figure}[hbt!]
    \centering
    \includegraphics[width=1\linewidth]{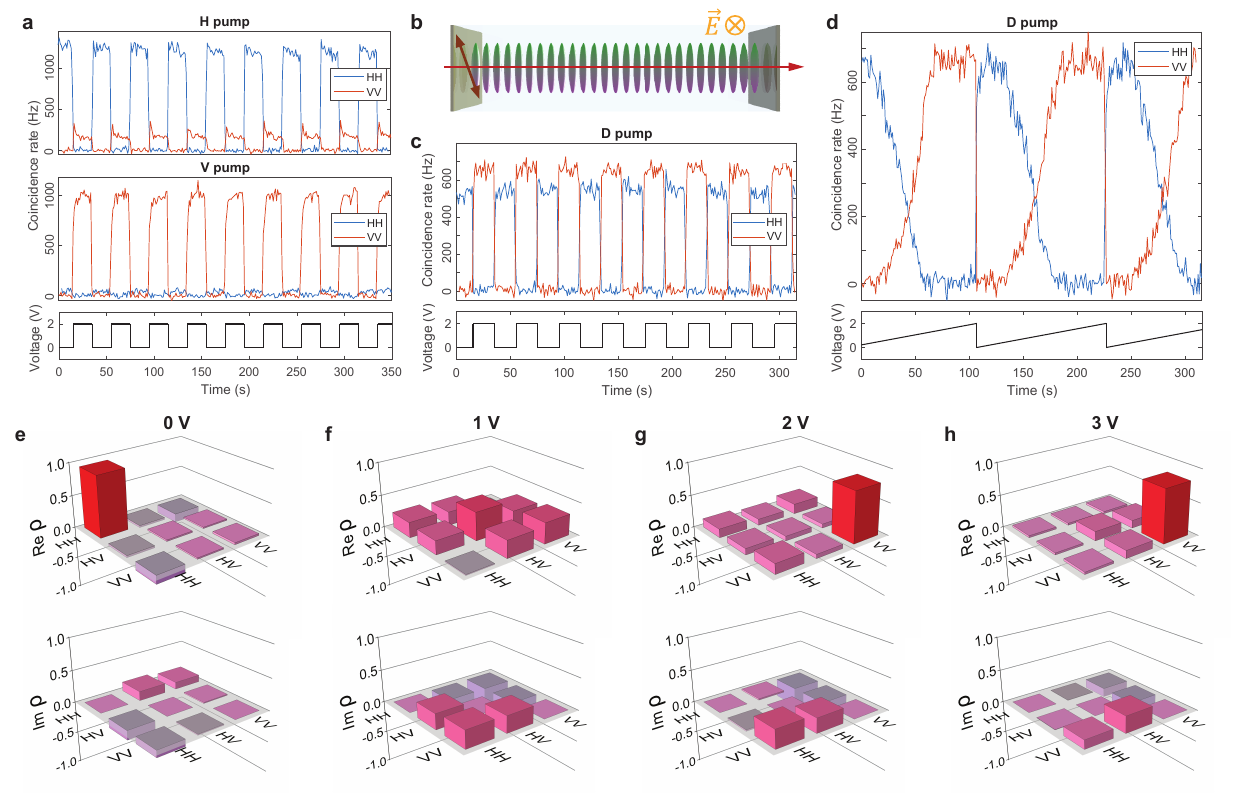}
    \caption{\textbf{Electric-field tuning of SPDC.} The electric field reorients the molecules, changing both the generation rate and the polarization state of photon pairs. {\bf{a)}} The rate of pair generation follows the electric field time dependence and can be modulated significantly for both horizontally (top) and vertically (bottom) polarized pump. The horizontally (blue) and vertically (orange) polarized states are turned on and off anti-phased. {\bf{b)}} and {\bf{c)}} The source can be switched between emitting horizontally and vertically polarized states with the diagonally polarized pump. In the absence of the field, the molecules are oriented horizontally. {\bf{d)}} The rate of pairs varied by applying a sawtooth voltage shows the existence of a threshold voltage required for reorienting the molecules, and the saturation voltage, where further increase brings no change. The polarization state also evolves gradually, according to the reconstructed polarization density matrices under \SI{0}{V} ({\bf{e}}), \SI{1}{V} ({\bf{f}}), \SI{2}{V} ({\bf{g}}) and \SI{3}{V} ({\bf{h}}).}
    \label{field}
\end{figure}

Next, we show how the generated two-photon state changes when the molecules are reoriented under the applied electric field. We measure the rate of real (non-accidental) coincidences for different pump polarizations and polarizations of photon pairs selected with two polarization filters before the detectors. With no field applied, horizontally (H-) oriented molecules generate H-polarized photon pairs from the H-polarized pump and no photon pairs from the vertically (V-) polarized pump (Fig.~\ref{field}a). Under a field applied perpendicular to the initial molecular orientation, the molecules align with the field (Fig.~\ref{overview}a, central and right panels), switching on photon pairs generation. Now, V-polarized photon pairs are generated from the V-polarized pump. The switching happens relatively fast, namely in $\approx$ \SI{0.5}{s}, as measured with SHG (Extended Data Fig. 4, see Methods).

The two-photon state switching is even more pronounced when photon pairs are generated with the diagonally (D-) polarized pump (Fig.~\ref{field}b). The coincidence rate traces are shown in Fig.~\ref{field}c. It follows that while the generation efficiency is defined by the overlap between the molecular orientation and the pump polarization, the generated state is solely defined by the molecular orientation. The state switching occurs gradually with the increase of the applied field (Fig.~\ref{field}d), starting at some threshold voltage and reaching the maximum at the saturation point, between \SIrange{1}{2}{V}, when practically all molecules are oriented along the field. Therefore, a voltage of only \SI{2}{V} across a \SI{500}{\micro m} gap is enough for almost complete switching of the generated two-photon state.

We further investigate the polarization of photon pairs by reconstructing the two-photon state. For classical light or a single photon, the polarization state is a superposition of two basis states, for instance, horizontal and vertical. In contrast, a two-photon state is described by four basis states~\cite{White2001}. However, in the case where two photons are distinguishable in no other way than polarization, the dimensionality is reduced to three, and the state is a qutrit~\cite{BurlakovPRA1999,Burlakov1999},
\begin{equation}
    \left|\Psi\right\rangle = C_1 \left|2\right\rangle_H\left|0\right\rangle_V + C_2  \left|1\right\rangle_H\left|1\right\rangle_V + C_3  \left|0\right\rangle_H\left|2\right\rangle_V,
    \label{qutrit}
\end{equation}
where $C_1$, $C_2$, and $C_3$ are complex amplitudes, so that $|C_1|^2+|C_2|^2+|C_3|^2=1$, and $\left|N\right\rangle_P$ is a Fock state with $N$ photons in polarization mode $P$. The corresponding density matrix $\rho$ extends the description to the case of mixed states. Since we detect photon pairs emitted into the same spatial collinear mode and do not distinguish them in frequencies, we characterize the two-photon state by a three-dimensional density matrix. We reconstruct the two-photon polarization density matrix via polarization tomography~\cite{Burlakov2003} optimized via maximum likelihood method~\cite{White2001}. The details of the experimental procedure can be found in Methods.

As we see in Fig. 2, e-h, as the electric field is gradually applied, the two-photon polarization state evolves from both photons polarized horizontally (e) through an intermediate state (f) to the state of both photons polarized vertically (g), which does not change significantly as the field is further increased (h). Therefore, we can obtain either two horizontally or vertically polarized photons or any intermediate two-photon polarization state with the same pump polarization by changing the molecular orientation via an applied electric field.

\section*{Engineering photon pairs via the sample geometry}
\begin{figure}[hbt!]
    \centering
    \includegraphics[width=0.85\linewidth]{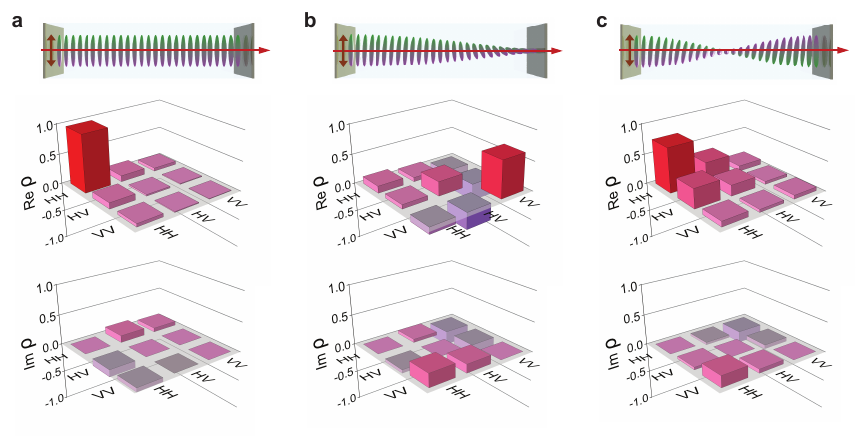}
    \caption{\textbf{Tuning SPDC by changing the twist geometry.} The density matrix of the two-photon polarization state strongly depends on the twist of the molecules along the sample. Examples of $0$ ({\bf{a}}), $\pi/2$ ({\bf{b}}) and $\pi$ ({\bf{c}}) twist angles were examined. The horizontal pump beam polarization coincides with the molecules' axes at the sample's start.}
    \label{twist}
\end{figure}

Finally, we investigate how the twist of molecular orientation along the sample affects the generated two-photon state (Fig.~\ref{twist}). As expected, a FNLC with no twist and horizontal orientation of the molecules generates pairs of H-polarized photons (Fig.~\ref{twist}a) from an H-polarized pump, according to the nonlinear tensor of the FNLC. In contrast, if molecules gradually change their orientation along the sample from horizontal to vertical ($\pi/2$-twist, Fig.~\ref{twist}b), the two-photon state is completely different: it contains mostly `VV' photon pairs with a small fraction of `HV' pairs. Further increase of the twist to $\pi$-twist brings the state close to the H-polarized two-photon state (Fig.~\ref{twist}c). Similar to the applied electric field, the molecular orientation twist gradually changes the state from two photons co-polarized along the molecular orientation (when there is no twist) to an orthogonal state in a superposition with cross-polarized photons.

\begin{figure}[hbt!]
    \centering
    \includegraphics[width=0.75\linewidth]{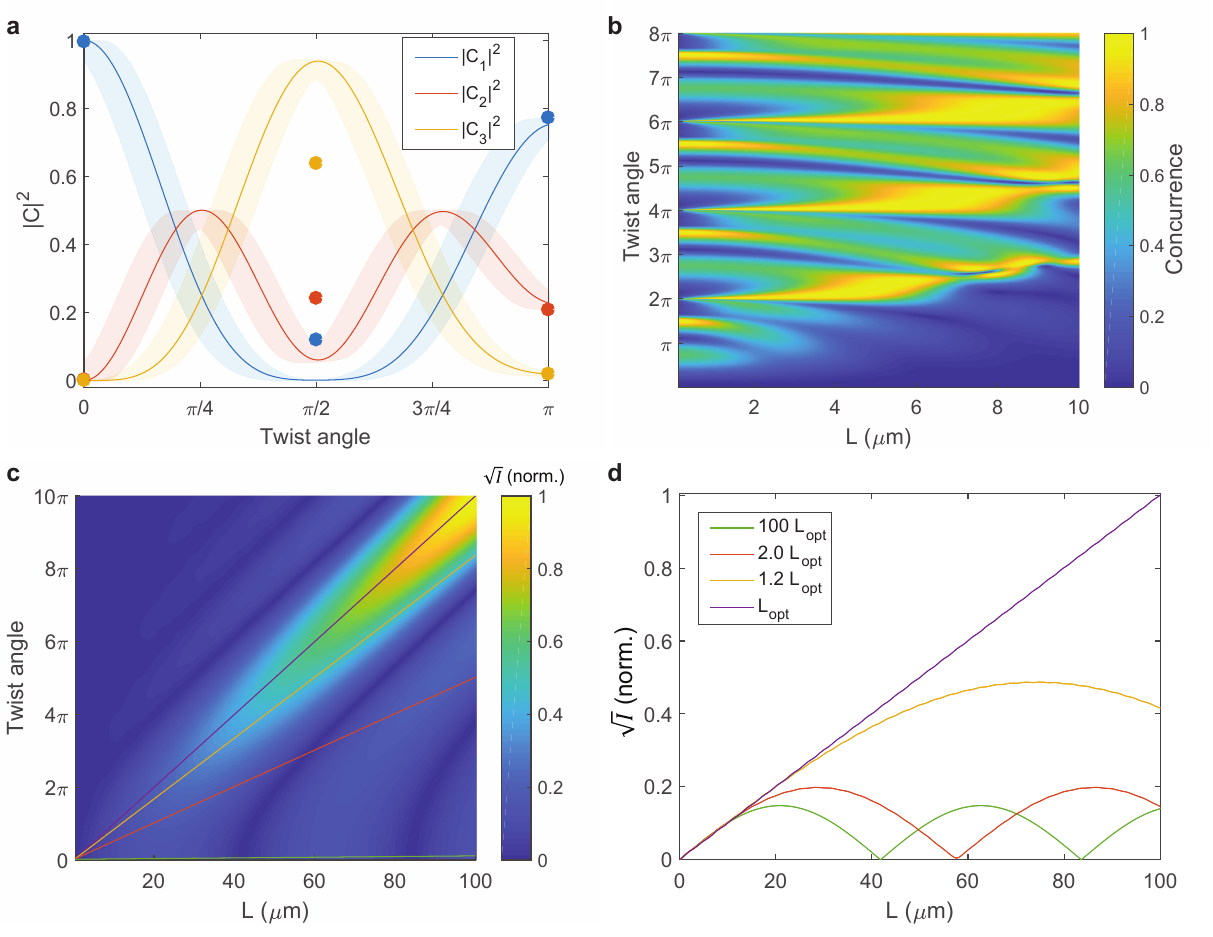}
    \caption{\textbf{Perspectives of SPDC engineering with molecules' twists.} 
    {\bf{a)}} Probabilities of the three basis two-photon states calculated as functions of the molecules' twist (lines) and measured for the three studied samples (points). 
    {\bf{b)}} Concurrence of the polarization state calculated versus the sample length and the molecules' twist.
    {\bf{c)}} Square root of the pair generation rate $I$ calculated versus the sample length and molecules' twist along the sample. Straight lines show the cases of different twist pitches. {\bf{d)}}
    Dependences of $\sqrt{I}$ on the sample length for various pitches. For an optimal pitch $L_{opt}=\SI{19.96}{\micro m}$, 
    the rate scales quadratically with the length, as in perfectly phase-matched crystals.}
    \label{theory}
\end{figure}

These results suggest that a broad range of two-photon polarization states can be achieved by engineering the source. To confirm it, we develop a theoretical model (see Methods) to investigate the effect of the parameters, such as the FNLC cell length and the molecular twist, on the photon pair generation. For the FNLC parameters investigated in the experiment (thickness \SIrange{7}{8}{\micro m} and twists 0, $\pi/2$, and $\pi$), our model has a fairly good agreement with the experimental results (Fig.~\ref{theory}a). The differences in the calculated and measured polarization states at \SI{90}{\degree} twist could be attributed to two things. Firstly, the actual twist could slightly differ from \SI{90}{\degree}, while, secondly, there is a possibility of slight differences between the linear twist considered for simulations and the actual twist profile of the average molecular orientation across the cell. These concerns were deduced from optical observations (Extended Data Fig.~1).

Further analysis of the parameter space shows that we can achieve two-photon polarization states that, after splitting the pair on a non-polarizing beamsplitter, yield pairs with any given degree of polarization entanglement. The concurrence of such a two-photon state $C=|2C_1C_3-C_2^2|$, a measure of polarization entanglement~\cite{Sultanov2022}, spans the whole range of values from 0 to 1 (Fig.~\ref{theory}b). 

Moreover, a proper twist of the molecular orientation along the FNLC slab can compensate for the nonlinear phase mismatch~\cite{zhao2022nontrivial}, as shown in Fig.~\ref{theory}c and d, similar to the periodic poling of a nonlinear crystal. The generation efficiency can be significantly enhanced by properly choosing the twist pitch along a macroscopically thick FNLC. For an optimal twist pitch, in our case $L_{opt}=\SI{19.96}{\micro m}$, which is very close to the coherence length $\SI{19.25}{\micro m}$, the rate scales quadratically with the length,  same as for SPDC in a phase-matched source. Experimentally, the proper pitch could be easily achieved by doping the FNLC with a chiral dopant \cite{nishikawaNFC2021,zhaoNFC2021,fengNFC2021}. In this case, a \SI{200}{\micro m} sample will generate 625 times more pairs than the current sample (Fig.~\ref{theory}d), reaching count rates of almost $1\, {\mathrm{MHz}}$, more than enough for most quantum technological applications. Such performance and the possibility to dynamically control the two-photon state are superior to existing crystal SPDC or fiber SFWM sources. The latter are less efficient, require a strong pulsed pump and a relatively long nonlinear medium, and have no polarization tunability without re-designing the source~\cite{kultavewuti2017, Takesue2008}.

In conclusion, we have demonstrated the first-ever successful generation of entangled photons via spontaneous parametric down-conversion in a liquid crystal, with an efficiency as high as of the most efficient commonly used nonlinear crystals of the same thickness. 

One of the most remarkable features discovered in these experiments is the unprecedented tunability of the two-photon state, achieved by manipulating the liquid crystal molecular orientation. By re-orienting the molecules through the application of an electric field, we can dynamically switch the polarization state of the generated photon pairs. This level of control over the photon pairs' polarization properties is a crucial advancement, offering novel opportunities for quantum state engineering in the sources with pixelwise-tunable optical properties, both linear and nonlinear.  

Alternatively, we can manipulate the polarization state by implementing a molecular orientation twist along the sample. This approach adds versatility to the design and utilization of liquid crystal-based photon pair sources. Moreover, a strong twist along the sample can dramatically increase the efficiency of a macroscopically large source, similar to the periodic poling of bulk crystals and waveguides, but much simpler technologically, since the structure is self-assembled and may be tuned with temperature and electric field~\cite{zhao2023high}. Due to their nonlinear coefficient comparable to the best nonlinear crystals, such as lithium niobate, and high damage threshold, FNLCs are perfectly suitable for practical applications. Furthermore, high-quality liquid crystal devices such as liquid crystal displays (LCDs) are made on an industrial scale, which, combined with our work, opens a path to scalable and cheap production of quantum light sources while exceeding the existing ones in efficiency and functionality.

In the future, the electric field tuning could be expanded to multi-pixel devices, which have the potential to generate tunable high-dimensional entanglement and multiphoton-states. Further, FNLCs can self-assemble in a variety of complex topological structures, which are expected to emit photon pairs in complex, spatially varying beams (structured light), such as vector and vortex beams~\cite{brasselet2009optical}. The liquid nature of FNLCs opens a path to their integration with existing optical platforms such as fibers~\cite{Du2004}, waveguides~\cite{Tripathi2020}, and metasurfaces~\cite{Izdebskaya2023}.

Overall, the results presented in this paper highlight the potential of liquid crystals for practical applications in quantum technologies. Liquid crystal-based photon-pair generation with tunable polarization states offers exciting possibilities for quantum information processing, quantum key distribution, and quantum-enhanced sensing.

\printbibliography

\section*{Methods}
\subsection*{Material and sample preparation}
The material employed in this study is ferroelectric nematic liquid crystal (FNLC) FNLC-1751 supplied by Merck Electronics KGaA. FNLC-1751 shows a stable ferroelectric nematic phase at room temperature, with the phase sequence Iso – $\SI{87}{\degree C}$ N – $\SI{57}{\degree C}$ - M2 - $\SI{45}{\degree C}$ – NF on cooling, where Iso refers to the isotropic phase, N to the non-polar nematic phase, M2 to the so-described splay modulated antiferroelectric nematic phase \cite{nishikawa2017fluid, FLC2, FLC3} and NF to the ferroelectric nematic phase. 

The material was confined in glass liquid crystal (LC) cells filled by capillary forces at $\SI{100}{\degree C}$ in the isotropic phase. We employed both commercially available and homemade cells. In the latter case, soda lime glass square $\SI{2}{cm}\times\SI{2}{cm}$ plates coated with transparent Indium Tin Oxide (ITO) conductive layer were assembled with plastic beads (EPOSTAR) spacers to achieve a variety of cells with different thicknesses ranging from \SI{7} to \SI{8}{\micro m}. In the bottom glass, ITO electrodes with a \SI{500}{\micro m} gap prepared by etching created the applied in-plane fields. Additionally, both plates were treated with a 30\% solution of polyimide SUNEVER 5291 (Nissan) film and rubbed to achieve orientational in-plane anchoring (planar alignment) of the liquid crystal. Combinations of different relative rubbing directions of the top and bottom glass plates ($0$, $\pi/2$ or $\pi$) result in different twist structures of the liquid crystal sample in the ferroelectric nematic phase \cite{chen2020first, Sebastian2021, FLC6}. In the three cases, the electrode glass was rubbed along the gap.  In-plane switching (IPS) LC cells purchased from Instec were employed for switching experiments on $\pi$-twisted structures. The cells had interdigitated electrodes in one of the substrates with alternating polarity; both the electrode width and the gap between them were \SI{15}{\micro m}. 
The surfaces had antiparallel rubbing along the electrodes (aligning agent KPI300B). 

After filling at $\SI{95}{\degree C}$, the sample was brought to room temperature by controllably decreasing the temperature at a rate of $\SI{0.5 }{\degree C/min}$. In the case of initially large domains breaking down into smaller ones after applying large voltages, the initial configuration was restored by reheating the samples and subsequent controlled cooling back to room temperature. The quality of the achieved alignment was inspected via polarizing optical microscopy and comparison with transmission spectra simulations (Berreman 4x4 matrix method performed with the open software package “dtmm” \cite{andrej_petelin_2020_4266242}, cell thickness \SI{7.6}{\micro m} and ordinary and extraordinary refractive indices as given in Extended Data Fig.~2).

\subsection*{Photon pair generation and detection}
The scheme of the experimental setup used for photon pair generation and detection is depicted in Extended Data Fig.~5. As a pump, we used a continuous-wave (CW) pigtailed single-mode fiber diode laser with a central wavelength of \SI{685}{nm}. After the power and polarization control, the pump beam was focused into the LC cell with the focusing spot size of \SI{5}{\micro m}. The maximum delivered pump power did not exceed \SI{10}{mW}. A function generator applied to the cell an electric field with different time profiles (see Fig. 2 of the main manuscript). The generated photon pairs were collected with a lens with a numerical aperture (NA) of 0.69. Then, a set of long-pass filters (LP) with the cut-on wavelength no longer than \SI{1250}{nm} cut the pump and short-wavelength photoluminescence from the sample and the optical elements of the setup. Photon pairs were further sent into a Hanbury Brown - Twiss - like setup comprising a non-polarizing beam splitter (NPBS) and two super-conducting nanowire single-photon detectors (SNSPDs). At each output of the NPBS, we placed a set of a half-wave plate (HWP), a quarter-wave plate (QWP), and a polarizing beam splitter (PBS), which acted as a polarization filter. The arrival time differences between the pulses of both SNSPDs were registered by a time-tagging device.

\subsection*{Two-photon spectrum measurement}
Since the SPDC radiation from an \SI{8}{\micro m} layer is extremely weak, the direct measurement of the two-photon spectrum (i.e., with a spectrometer or optical spectrum analyzer) is nearly impossible. Therefore, we measured the spectrum of the detected photon pairs via single-photon fiber spectroscopy~\cite{Valencia2002}. Before one of the SNSPDs, we inserted a \SI{2}{km} long dispersion-shifted fiber with the zero-dispersion wavelength at \SI{1.68}{\micro m}. Due to the dispersion of the fiber, the photon wave-packet stretched in time, resulting in a spread of the coincidence peak, which then inherited the spectrum's features and the spectral losses of the setup. We acquired the coincidence histogram with different sets of spectral filters (Extended Data Fig.~6, a and b) to map the arrival time differences to the corresponding wavelengths of the dispersed photon. The calibration curve (Extended Data Fig.~6b) was obtained by fitting the reference points with a quadratic polynomial function. Yet, the spectrum is strongly affected by the spectral losses of the setup and the dispersive fiber. For that reason, we additionally measured the spectrum of photon pairs generated in a thin (\SI{7}{\micro m}) layer of LiNbO$_3$ (Extended Data Fig.~6c), where the generated two-photon spectrum is mostly flat, up to a modulation by the Fabry-Perot effect inside the layer. We then used the spectrum of photon pairs from the LiNbO$_3$ wafer as a reference spectrum. 

\subsection*{Two-photon state reconstruction}
We performed quantum tomography to reconstruct the two-photon polarization state generated in the LC. The procedure is analogous to measuring the Stokes parameters for classical light or a single photon. By measuring the pair detection rates for different polarization states filtered in the two arms of the HBT setup, we were able to reconstruct the density matrix of the two-photon state. Since there is no prior assumption about the generated two-photon state, we performed all 9 required measurements for the reconstruction of the $3\times 3$ density matrix. The full protocol is described in Extended Data Table~1. The values in the table refer to the orientation of the fast axis of each waveplate with respect to the horizontal direction. It is worth mentioning that the described protocol does not take into account the mirroring effect of polarization in the reflected arm of the HBT setup. Therefore, either the angles of the wave plates in the reflected arm must be changed to the opposite values, or an odd number of mirrors must be used in the reflected arm of the HBT setup. The protocol used for the qutrit state reconstruction is the reduced version of the protocol for the reconstruction of the two-photon polarization state with two distinguishable photons (ququart state)~\cite{White2001}.

To avoid systematic errors in the density-matrix reconstruction, we additionally post-processed the measured data using the maximum-likelihood method (MaxLi). MaxLi aims to find the density matrix, closest to the measured one, that satisfies all basic physical properties of a density matrix. We used a procedure similar to the one described in Ref.~\cite{White2001} with minor modifications (see Section 4 of Supplementary Information).

\subsection*{SPDC in liquid crystals - theoretical model}
We developed a theoretical model to predict the polarization two-photon state generated via SPDC in a nonlinear LC with an arbitrary but linear molecular orientation twist along the cell. The goal is to determine the complex amplitudes of the polarization two-photon state $C_1$, $C_2$, and $C_3$ from Eq. (\ref{qutrit}) of the manuscript. We assumed a single-mode, collinear, and frequency-degenerate photon pair generation in the plane wave approximation for simplicity. However, the model can be further extended towards the multi-mode regime of SPDC with realistic angular and frequency spectra, as well as for the case of a non-gradual molecular twist.

Due to weak interaction, we can use the perturbation theory for the unitary transformation of the state vector~\cite{Loudon}. The state can be written as
\begin{equation}
    \vert\Psi\rangle = \vert vac\rangle+C\int\limits_{-L}^{0}dz\hat{\chi}^{(2)}(z)\vdots \vec{e_s}^*(z)\vec{e_i}^*(z)\vec{e}_p(z) a^{\dagger}_sa^{\dagger}_i\vert vac\rangle,
    \label{integral}
\end{equation}
where $a^{\dagger}_s,a^{\dagger}_i$ are the photon creation operators for signal and idler photons, each of them defined in some polarization eigenmode, and the polarization vectors $\vec{e}_{s,i,p}(z)$ also encode the phase accumulation during the propagation along the crystal. The constant $C$ contains only the information about the overall generation efficiency and, therefore, is of no interest to us.

For convenience, however, we use two polarization bases instead of the polarization eigenmodes. The first basis is a standard linear polarization basis with horizontal and vertical polarizations determined with respect to the laboratory coordinate system, $\{H,V\}$. In this basis, the two-photon polarization state can be expressed as a qutrit state (\ref{qutrit}) since two photons are assumed to be indistinguishable in all other Hilbert spaces apart from polarization~\cite{Burlakov1999}. The final goal of the calculations is to determine complex amplitudes $C_{1,2,3}$ from Eq.~(\ref{qutrit}). Since the molecular orientation changes along the crystal and implies the spatial modulation of the nonlinearity, it is more convenient to calculate the convolution of the $\chi^{(2)}$ tensor with the polarization vectors of the interacting photons in the second basis aligned with the instant orientation of the molecules, $\{e, o\}$. We denote the corresponding projections with indices $e$ and $o$ for the linear polarization along and orthogonal to the instant molecular orientation, respectively. Instead of $\chi^{\left(2\right)}$ tensor, we use the standard notation of Kleinman $d$ tensor. Therefore, the convolution is written as
\begin{multline}
\chi^{\left(2\right)}\vec{e_s}^*\vec{e_i}^*\vec{e}_p = {e_s^o}^*{e_i^o}^*\left(d_{22}\,e_p^o+d_{32}\,e_p^e\right) + {e_s^e}^*{e_i^e}^*\left(d_{23}\,e_p^o+d_{33}\,e_p^e\right) + \\ + \left({e_s^e}^*{e_i^o}^*+{e_s^o}^*{e_i^e}^*\right)\left(d_{24}\,e_p^o+d_{34}\,e_p^e\right)\,,
\label{conv}
\end{multline}
where the polarization basis vectors and the tensor components are functions of $z$, and the convolution is defined in the local coordinate system of the molecules. The z-direction is defined in the same way for both bases and denotes the photon propagation direction along the crystal.

To calculate the polarization two-photon state, we consider a liquid crystal with a uniform rotation of the molecules along the crystal (Extended Data Fig.~7). At an arbitrarily chosen layer of thickness dz at position z, the pump polarization is modified by all the previous layers it has passed through. The polarization state of photon pairs generated from the corresponding layer dz is further modified by all subsequent layers of the liquid crystal. The final state at the output of the crystal is the superposition of all polarizations generated along the crystal. Therefore, to calculate the output two-photon polarization state, we integrate the contribution of each layer of the liquid crystal taking into account the corresponding polarization transformations of both the pump and the incremental photon pair state generated from each layer.

To calculate the propagation of the pump, the initial pump polarization is represented by a Jones vector (Extended Data Fig.~7) in the $\{H,V\}$ basis. The angle $\varphi_0$ is defined as the angle between both coordinate systems at the beginning of the sample, i.e. the angle between the global coordinate $H$ direction and the extraordinary molecule axis $e$ at the beginning of the sample. The first step is to bring the pump from the global basis to the local basis at the beginning of the sample via rotating the pump Jones vector by $\varphi_0$:
\begin{equation}
\vec{e_p}^{in}=R\left(\varphi_0\right)\,\vec{e_p}^0\,,
\end{equation}
where R is the standard rotation matrix. The polarization transformation of light propagating through a twisted nematic liquid crystal with a uniform twist is described by the corresponding Jones matrix~\cite{yariv1983optical, moreno1998twist, lu1990theory},
\begin{equation}
M_{TLC}=e^{i\phi}\,R\left(-\varphi\right)\,M\left(\varphi, \beta\right)\,; \qquad 
M\left(\varphi, \beta\right)=
\begin{pmatrix}
\cos{X}+i{\frac{\beta}{X}}\sin{X} & {\frac{\varphi}{X}}\sin{X}\\
-{\frac{\varphi}{X}}\sin{X} & \cos{X}-i{\frac{\beta}{X}}\sin{X}
\end{pmatrix},
\label{M_TLC}
\end{equation}
where $\phi=\tilde{k}l$ with $\tilde{k}={\frac{1}{2}}\left(k^e+k^o\right)$ being the average $k$ vector, $l$ is the sample length, $\varphi$ is the twist angle, and $\beta=\pi l\left(n_e-n_o\right)/\lambda=gl$ characterizes birefringence, where $g={\frac{1}{2}}\left(k^e-k^o\right)$. Additional parameter $X$ is defined as $X=\sqrt{\varphi^2+\beta^2}$.

At a certain chosen position z, the pump polarization is transformed by the part of the liquid crystal from $-L$ to $z$, with the effective length of this layer being $z+L$. The pump polarization vector in the local basis at position $z$ then has the form
\begin{equation}
\begin{pmatrix}
e_p^e \left(z\right)\\
e_p^o \left(z\right)
\end{pmatrix}
=e^{i\tilde{k}_p\left(z+L\right)}M\left({\tfrac{z+L}{L}}\varphi, \left(z+L\right)\!g_p\right)R\left(\varphi_0\right)
\begin{pmatrix}
e_p^{\scriptscriptstyle H}\\
e_p^{\scriptscriptstyle V}
\end{pmatrix},
\end{equation}
where $\varphi$ denotes the full twist of the sample. We intentionally leave the pump polarization defined in the local basis as it is convenient for calculating its convolution with $\hat{\chi}^{(2)}$. We explicitly write the pump polarization vector at position $z$ in the local basis as a function of the input pump polarization in the $\{H,V\}$ basis,
\begin{align}
e_p^e \left(z\right) &=\left[t_p\,e_p^{\scriptscriptstyle H}+r_p\,e_p^{\scriptscriptstyle V}\right]e^{i\tilde{k}_p\left(z+L\right)}\notag,\\
e_p^o \left(z\right) &=\left[t_p^*\,e_p^{\scriptscriptstyle V}-r_p^*\,e_p^{\scriptscriptstyle H}\right]e^{i\tilde{k}_p\left(z+L\right)}\,,
\label{transformation1}
\end{align}
where
\begin{align}
t_p&=\xi_p\cos{\varphi_0}-\mu_p\sin{\varphi_0}\notag,\\
r_p&=\xi_p\sin{\varphi_0}+\mu_p\cos{\varphi_0}\notag,\\
\xi_p&=\cos{\left({\tfrac{z+L}{L}}X_p\right)}+i{\frac{g_pL}{X_p}}\sin{\left({\tfrac{z+L}{L}}X_p\right)},\\
\mu_p&={\frac{\varphi}{X_p}}\sin{\left({\tfrac{z+L}{L}}X_p\right)}\notag,\\
X_p&=\sqrt{\varphi^2+\left(g_pL\right)^2}\notag.
\end{align} 

By inserting these expressions into (\ref{conv}), we can find the polarization state of photon pairs generated from a unit layer at position $z$ in the local basis. However, since we are interested in the output polarization state, the polarization of both signal and idler photons must be propagated from $z$ to the end of the crystal in a similar way. This transformation can be written as
\begin{equation}
\begin{pmatrix}
e_{s,i}^{\scriptscriptstyle H} \\
e_{s,i}^{\scriptscriptstyle V} 
\end{pmatrix}
=R\left(-\varphi_0-\varphi\right)e^{i\tilde{k}_{s,i}\left(-z\right)}M\left({\tfrac{-z}{L}}\varphi, -zg_{s,i}\right)
\begin{pmatrix}
e_{s,i}^e\left(z\right)\\
e_{s,i}^o\left(z\right)
\end{pmatrix},
\end{equation}
where the photons are propagating from $z$ to $0$. The explicit form of the output polarization for the signal and idler photons generated at $z$ is
\begin{align}
e_{s,i}^{\scriptscriptstyle H} &=\left[t_{s,i}\,e_{s,i}^e\left(z\right)+r_{s,i}\,e_{s,i}^o\left(z\right)\right]e^{-i\tilde{k}_{s,i}z}\notag\\
e_{s,i}^{\scriptscriptstyle V} &=\left[t_{s,i}^*\,e_{s,i}^o\left(z\right)-r_{s,i}^*\,e_{s,i}^e\left(z\right)\right]e^{-i\tilde{k}_{s,i}z}\,,
\label{transformation2}
\end{align}
with similar notation as before,
\begin{align}
t_{s,i}&=\xi_{s,i}\cos{\left(\varphi_0+\varphi\right)}+\mu_{s,i}^*\sin{\left(\varphi_0+\varphi\right)}\notag\\
r_{s,i}&=-\xi_{s,i}^*\sin{\left(\varphi_0+\varphi\right)}+\mu_{s,i}\cos{\left(\varphi_0+\varphi\right)}\notag\\
\xi_{s,i}&=\cos{\left({\tfrac{-z}{L}}X_{s,i}\right)}+i{\frac{g_{s,i}L}{X_{s,i}}}\sin{\left({\tfrac{-z}{L}}X_{s,i}\right)}\\
\mu_p&={\frac{\varphi}{X_{s,i}}}\sin{\left({\tfrac{-z}{L}}X_{s,i}\right)}\notag\\
X_{s,i}&=\sqrt{\varphi^2+\left(g_{s,i}L\right)^2} \,.\notag
\end{align} 
To perform convolution (\ref{conv}), Eq.~(\ref{transformation2}) needs to be reversed to express $e_{s,i}^{e,o}\left(z\right)$ as functions of the outcome polarizations $e_{s,i}^{\scriptscriptstyle H,V}$. With this transformation, alongside Eqs.~(\ref{conv}) and (\ref{transformation1}) the convolution is written as
\begin{equation}
\begin{split}
\chi^{\left(2\right)}\vec{e_s}^*\vec{e_i}^*\vec{e}_p = &\Bigl[\left(r_{s}\,{e_{s}^{\scriptscriptstyle H}}^*+t_{s}^*\,{e_{s}^{\scriptscriptstyle V}}^*\right)\left(r_{i}\,{e_{i}^{\scriptscriptstyle H}}^*+t_{i}^*\,{e_{i}^{\scriptscriptstyle V}}^*\right)P_1 + \left(t_{s}\,{e_{s}^{\scriptscriptstyle H}}^*-r_{s}^*\,{e_{s}^{\scriptscriptstyle V}}^*\right)\left(t_{i}\,{e_{i}^{\scriptscriptstyle H}}^*-r_{i}^*\,{e_{i}^{\scriptscriptstyle V}}^*\right)P_2  \Bigr.\\ +&\,\,\Bigl. \left(t_{s}\,{e_{s}^{\scriptscriptstyle H}}^*-r_{s}^*\,{e_{s}^{\scriptscriptstyle V}}^*\right)\left(r_{i}\,{e_{i}^{\scriptscriptstyle H}}^*+t_{i}^*\,{e_{i}^{\scriptscriptstyle V}}^*\right)P_3 + \left(r_{s}\,{e_{s}^{\scriptscriptstyle H}}^*+t_{s}^*\,{e_{s}^{\scriptscriptstyle V}}^*\right)\left(t_{i}\,{e_{i}^{\scriptscriptstyle H}}^*-r_{i}^*\,{e_{i}^{\scriptscriptstyle V}}^*\right)P_3 \Bigr] \\ &e^{i\tilde{k}_{p}L}e^{i\left[\tilde{k}_{p}-\left(\tilde{k}_{s}+\tilde{k}_{i}\right)\right]z}
\end{split},
\label{convfin}
\end{equation}
where further notation shortening was introduced via
\begin{align}
P_1&=d_{22}\left(t_p^*\,e_p^{\scriptscriptstyle V}-r_p^*\,e_p^{\scriptscriptstyle H}\right)+d_{32}\left(t_p\,e_p^{\scriptscriptstyle H}+r_p\,e_p^{\scriptscriptstyle V}\right)\notag,\\
P_2&=d_{23}\left(t_p^*\,e_p^{\scriptscriptstyle V}-r_p^*\,e_p^{\scriptscriptstyle H}\right)+d_{33}\left(t_p\,e_p^{\scriptscriptstyle H}+r_p\,e_p^{\scriptscriptstyle V}\right)\notag,\\
P_3&=d_{24}\left(t_p^*\,e_p^{\scriptscriptstyle V}-r_p^*\,e_p^{\scriptscriptstyle H}\right)+d_{34}\left(t_p\,e_p^{\scriptscriptstyle H}+r_p\,e_p^{\scriptscriptstyle V}\right)\,.
\end{align}
To find the state, we have to substitute the components of the Jones vectors $e^{H,V}_{s,i}$ with the corresponding photon creation operators. In this case, transformations (\ref{transformation1}, \ref{transformation2}) are equivalent to the unitary transformations of a beam splitter with two input and two output polarization modes. Substituting (\ref{convfin}) into (\ref{integral}) and grouping the components with the same pair of the creation operators, we can finally find the two-photon polarization state in the qutrit form (\ref{qutrit}) with the complex amplitudes
\begin{align}
c_1&=\sqrt{2}\int\displaylimits_{-L}^0{\rm{d}}z\Bigl[r_sr_iP_1+t_st_iP_2+\left(t_sr_i+r_st_i\right)P_3\Bigr]e^{i\left[\tilde{k}_{p}-\left(\tilde{k}_{s}+\tilde{k}_{i}\right)\right]z},\notag\\
c_2&=\int\displaylimits_{-L}^0{\rm{d}}z\Bigl[\left(r_st_i^*+t_s^*r_i\right)P_1-\left(t_sr_i^*+r_s^*t_i\right)P_2+\left(t_st_i^*-r_sr_i^*+t_s^*t_i-r_s^*r_i\right)P_3\Bigr]e^{i\left[\tilde{k}_{p}-\left(\tilde{k}_{s}+\tilde{k}_{i}\right)\right]z},\notag\\
c_3&=\sqrt{2}\int\displaylimits_{-L}^0{\rm{d}}z\Bigl[t_s^*t_i^*P_1+r_s^*r_i^*P_2-\left(t_s^*r_i^*+r_s^*t_i^*\right)P_3\Bigr]e^{i\left[\tilde{k}_{p}-\left(\tilde{k}_{s}+\tilde{k}_{i}\right)\right]z}.
\end{align}

The polarization state vector has to be further normalized with the norm $\sqrt{\vert C_1\vert^2+\vert C_2\vert^2+\vert C_3\vert^2}$. While we use the normalized values of the complex amplitudes for the analysis of the two-photon polarization state (Extended Data Fig.~8), the norm itself shows the relative generation efficiency for different parameters of the liquid crystal, such as length and twist (insets c and d in Fig. 4 of the main manuscript).

Further development of the model involves more strict quantum-optical calculations, with the real angular and spectral distributions of the generated photons, as well as the spatial properties of the pump beam, internal reflections of both the pump and the generated photons, etc. Furthermore, the approximation of a non-depleted pump is valid only in the low-gain regime of SPDC, while such a source is incredibly promising for generating squeezed vacuum and twin beams. Finally, we assume a perfect uniform twist of the molecules, which is hard to achieve experimentally, especially for twists not multiple to $\pi$. Although this model is significantly simplified, it proved to be reliable and provides a great insight into the physics of this type of material. 

\section*{End Notes}
\noindent{\textbf{Acknowledgements.}} The authors thank Merck Electronics KGaA for providing the FNLC material, Natan Osterman for suggesting the use of FNLC as the nonlinear medium, and Janja Milivojević for assembling LC cells.

\noindent{\textbf{Funding.}} The authors acknowledge financial support from the European Research Council (ERC) under the European Union’s Horizon 2020 research and innovation programme (grant agreement No. 851143), from Slovenian Research and Innovation Agency (ARIS) (P1-0099, P1-0192), from Deutsche Forschungsgemeinschaft (429529648 – TRR 306 QuCoLiMa). The project/research is part of the Munich Quantum Valley, which is supported by the Bavarian state government with funds from the Hightech Agenda Bavaria. V. S. and M. V. C. are part of the Max Planck School of Photonics supported by BMBF, Max Planck Society, and Fraunhofer Society.

\appendix
\renewcommand{\thefigure}{\arabic{figure}}
\renewcommand{\figurename}{Extended data Fig.}
\renewcommand{\thetable}{\arabic{table}}
\renewcommand{\tablename}{Extended data Table}
\setcounter{figure}{0}
\setcounter{table}{0}
\newpage
\begin{figure}[hbt!]
\centering
\includegraphics[width=10cm]{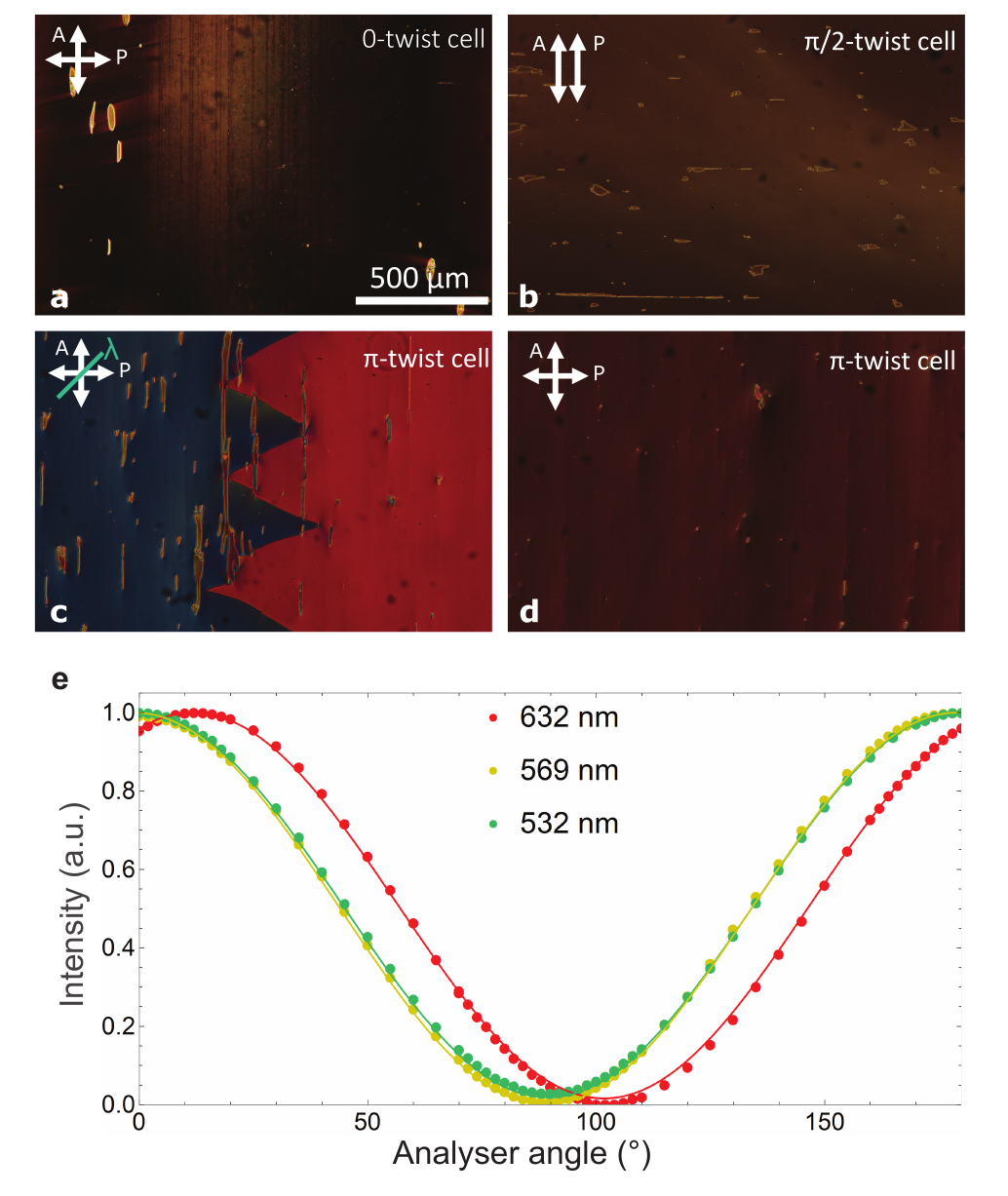}
\caption{{\bf{Polarization microscopy (POM) images of different 0-, $\pi$/2- and $\pi$-twisted cells.}} {\bf{a)}} A 0-twist cell as observed between crossed polarizers shows almost perfect extinction indicating uniform molecular orientation across the cell thickness. {\bf{b)}} $\pi$/2-twist cells as observed between parallel polarizers. The imperfect extinction indicates a slight deviation of the molecular orientation direction across the cell from the ideal $\pi$/2 linear twist. Molecular orientation direction across the cell is analysed in {\bf{d}}. {\bf{c)}} A $\pi$-twisted cell as observed between crossed polarizers and a full-wave plate $\lambda$ shows the division into two domains of opposite handedness (as evidenced by the sierra-walls), typical for antiparallel rubbed cells in which molecular orientation follows a $\pi$-twist structure across the sample thickness \cite{chen2021polar, Sebastian2021}. Elongated domains indicate surface defects, which are absent in the area of the electrode gap ({\bf{d}}), where a single domain was imaged. {\bf{e)}} Normalized transmitted intensity for a $\pi$/2-twist cell at different wavelengths (\SI{532}{nm}, \SI{569}{nm} and \SI{632}{nm}) as a function of the analyzer rotation where zero angle corresponds to the analyzer perpendicular to the polarizer. Circles denote experimental data acquired with the incoming polarization parallel to the bottom glass surface rubbing. Full lines correspond to transmission spectra simulations performed considering a linear twist structure for the molecular orientation. The best fits were obtained using \SI{84}{\degree} twist for \SI{532}{nm} and \SI{632}{nm} curves (with rotation from \SI{5}{\degree} to \SI{89}{\degree} from bottom to top surface, with respect to incoming polarisation) and \SI{85}{\degree} for \SI{569}{nm} curve (rotation from \SI{0}{\degree} to \SI{85}{\degree}). The fact that no simple linear twist profile was found that simultaneously fits the three sets of data, indicates that the twist profile could deviate slightly from linear. This variation, together with the twist not being exactly \SI{90}{\degree}, could explain the differences in the calculated and measured polarization state in Fig.~\ref{theory}a at $\pi/2$ twist.}
\label{ED1_POM}
\end{figure}

\newpage
\begin{figure}[hbt!]
\centering
\includegraphics[width=16cm]{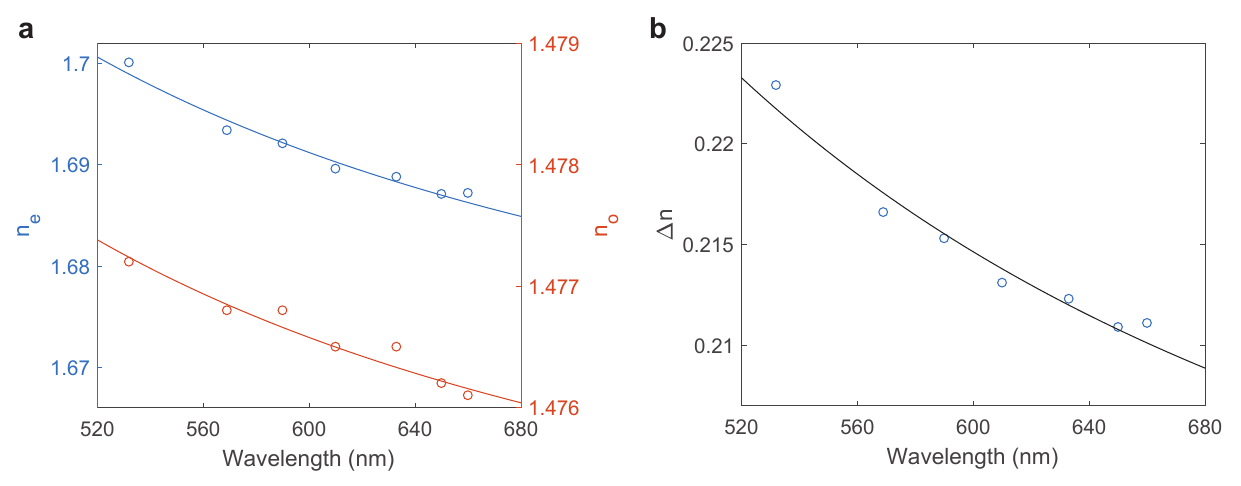}
\caption{{\bf{Refractive indices and birefringence of FNLC-1751 measured versus the wavelength.}} {\bf{a)}} Measured values and fitting functions for the extraordinary (blue) and ordinary (orange) refractive indices at different wavelengths. {\bf{b)}} Values of birefringence extracted from data in panel {\bf{a}}. In all cases, the fitting functions are two-term Cauchy models ($A+B/\lambda^2$). The resulting parameters are $A=1.663$, $B=\SI{10256}{nm^2}$ for the extraordinary refractive index and $A=1.474$, $B=\SI{874}{nm^2}$ for the ordinary refractive index.}
\label{ED2_ref_ind}
\end{figure}

\begin{figure}[hbt!]
\centering
\includegraphics[width=16cm]{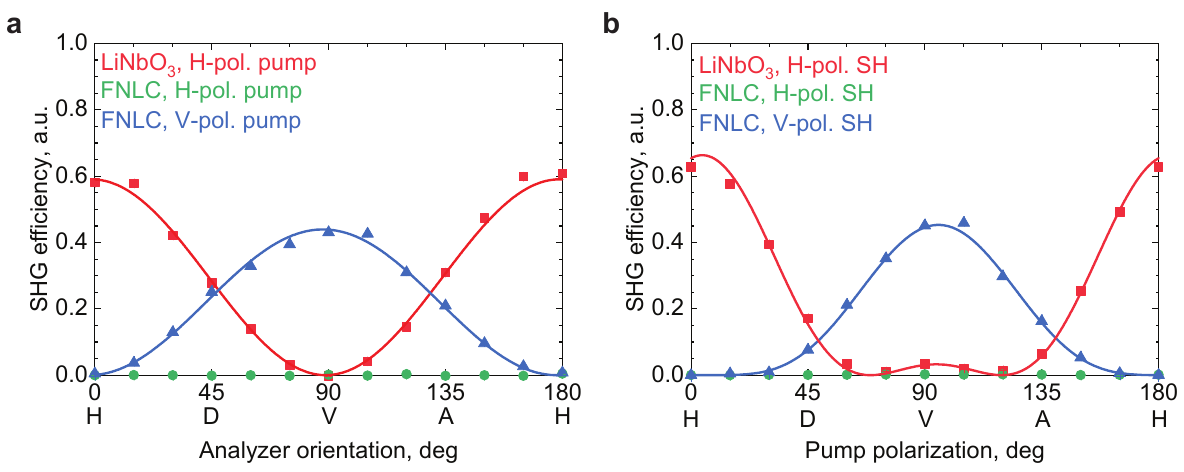}
\caption{{\bf{Measurement of the SHG efficiency in LC to determine the nonlinear tensor $d$.}} {\textbf{a)}} We measured the SHG efficiency with a fixed pump polarization (horizontal or vertical) in LC (green and blue points, respectively) and compared it with the SHG efficiency in LN (red points). {\textbf{b)}} Similar measurements were done with the fixed orientation of the analyzer and a variable pump polarization. After fitting the results with the corresponding theoretical curves and comparing the SHG efficiencies in LN and LC, we concluded that out of six components of the $d$ nonlinear tensor of LC that we were able to retrieve, only one is significant, $d_{33} \approx \SI{20}{pm/V}$.}
\label{ED3_chi2}
\end{figure}

\begin{figure}[hbt!]
\centering
\includegraphics[width=16cm]{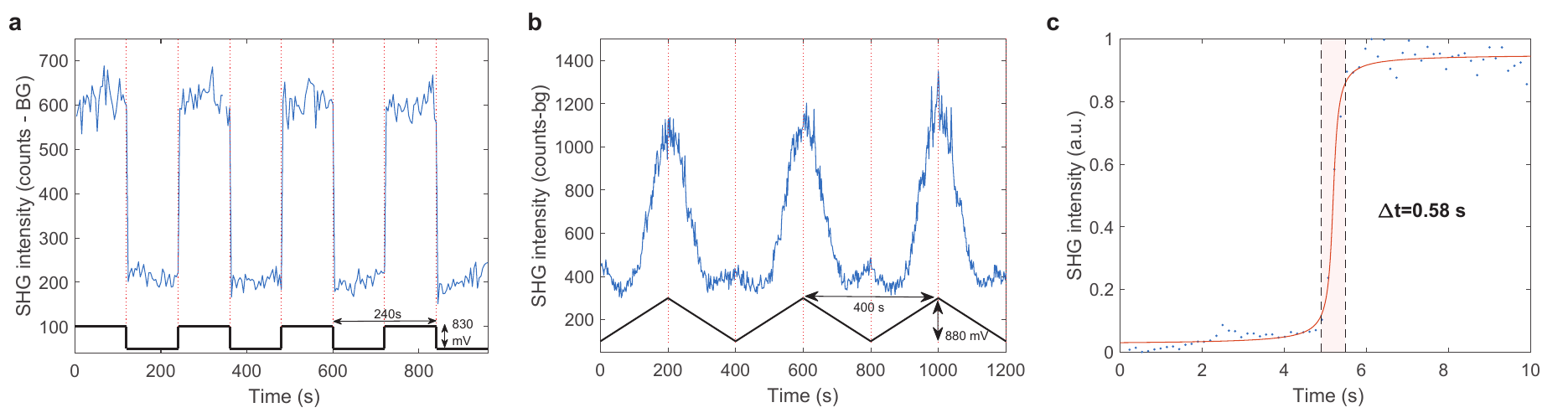}
\caption{{\bf{SHG switching under the applied electric field in the $\pi$-twisted LC.}} {\bf{a)}} Background-subtracted SHG intensity over time while applying square \SI{830}{mV} voltage with \SI{240}{s} period. We can see discrete transitions when the field is turned on or off and a stable behavior during the time the field remains fixed. The contrast between the on and off states of the field corresponds to a threefold increase in intensity. {\bf{b)}}  Background-subtracted SHG intensity over time while applying linearly increasing/decreasing \SI{880}{mV} voltage pulses. The SHG intensity follows the voltage profile, the only exception being the low-voltage regions where the intensity is not really changing within the measurement error. We attribute this to the fact that electric field is insufficient to reorient the molecules and change the sample structure. The contrast between the on and off state is again threefold. {\bf{c)}} Characteristic profile of the LC SHG intensity when the voltage is quickly switched on. The transition time of the molecular reorientation and relaxation into the new configuration driven by the applied field is around half a second (denoted by the shaded area). It is defined as the difference between the points where the amplitude is at 10\% and 90\%, respectively. The function fitted to the data is $a\cdot{\mathrm{atan}}\left(\left(x-c\right)/b\right)+d$. In the presented case, \SI{1}{V} was applied to the sample without the twist. No significant difference was observed for different field strengths or sample configurations (twist). The same behavior is expected also for SPDC, but it is harder to reliably measure with such temporal resolution due to much lower intensity.}
\label{ED4_SHG}
\end{figure}

\begin{figure}[hbt!]
\centering
\includegraphics[width=16cm]{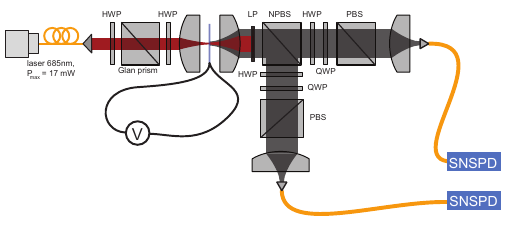}
\caption{{\bf{Experimental setup for generating and detecting entangled photons from LC.}} After the power and polarization control, the pump (CW laser light @ \SI{685}{nm}) was focused in the LC cell. Generated photon pairs were sent to the Hanbury Brown - Twiss setup and detected via coincidence measurements. HWP, half-wave plate; QWP, quarter-wave plate; LP, long-pass filter; NPBS, non-polarizing beam-splitter; PBS, polarizing beam-splitter; SNSPD, superconducting nanowire single-photon detector.}
\label{ED5_setup}
\end{figure}

\begin{figure}[hbt!]
\centering
\includegraphics[width=16cm]{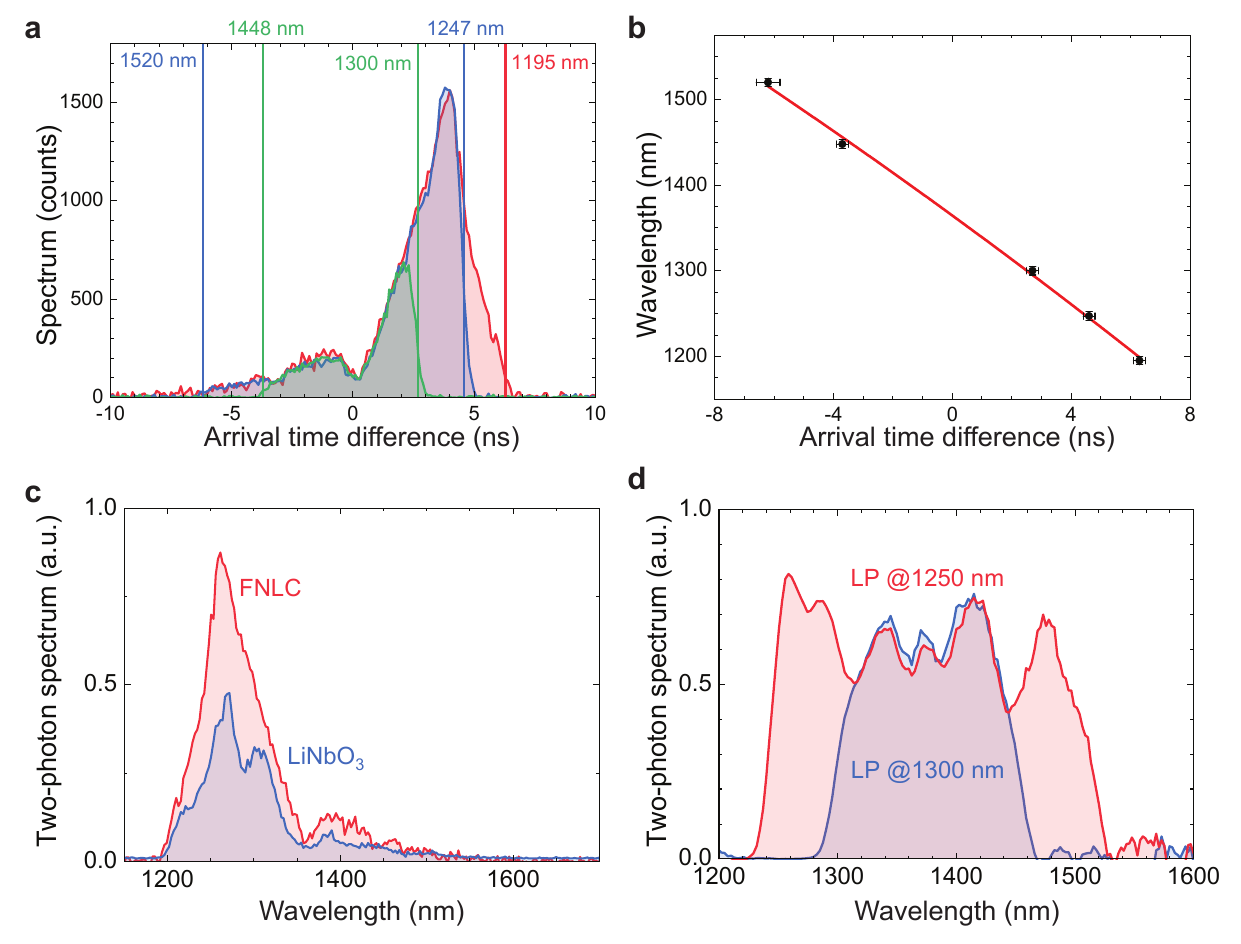}
\caption{{\bf{Two-photon spectrum measurement.}} {\bf{a)}} We measured the histogram of arrival time differences for several sets of frequency filters and matched their edges with the transmission spectra of the filters. {\bf{b)}} The obtained points were fitted to map the detection time delay to the wavelength of the dispersed photon. {\bf{c)}} We additionally measured the spectrum of \SI{7}{\micro m} thick LN wafer with a known two-photon spectrum as a reference to account for the spectral losses in the setup. {\bf{d)}} The final spectrum of photon pairs generated in LC was normalized to the reference spectrum. The measured spectrum is solely limited by the detection efficiency and the spectral filters used in the experiment. Due to the relaxed phase-matching, the generated two-photon spectrum should be much broader, occupying several octaves.}
\label{ED6_spectrum}
\end{figure}

\begin{table}[hbt!]
    \centering
    \begin{tabular}{|c|c|c|c|c|}
    \hline
    Polarization state & HWP$_1$, deg & QWP$_1$, deg & HWP$_2$, deg & QWP$_2$, deg\\
    \hline
    H-H & $0$ & $0$ & $0$ & $0$ \\
    H-V & $0$ & $0$ & $45$ & $0$ \\
    V-V & $45$ & $0$ & $45$ & $0$ \\
    H-D & $0$ & $0$ & $22.5$ & $45$ \\
    H-R & $0$ & $0$ & $0$ & $-45$ \\
    V-A & $45$ & $0$ & $-22.5$ & $45$ \\
    V-L & $45$ & $0$ & $0$ & $45$ \\
    D-D & $22.5$ & $45$ & $22.5$ & $45$ \\
    D-R & $22.5$ & $45$ & $0$ & $-45$ \\
    \hline
    \end{tabular}
    \caption{{\bf{The protocol for two-photon polarization tomography of a qutrit state.}} Indices 1 and 2 correspond to the different outputs of the NPBS.}
    \label{ED7_PT}
\end{table}

\begin{figure}[hbt!]
\centering
\includegraphics[width=15cm]{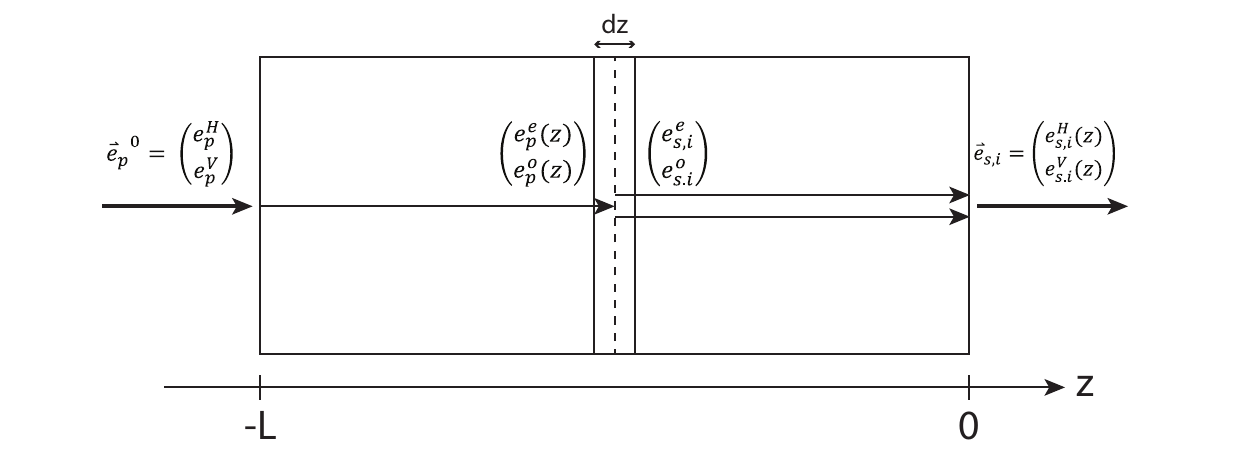}
\caption{\bf{Notation used to describe the propagation of the pump and the generated photon pairs within the sample in the theoretical model.}}
\label{ED8_skica}
\end{figure}

\begin{figure}[hbt!]
\centering
\includegraphics[width=16cm]{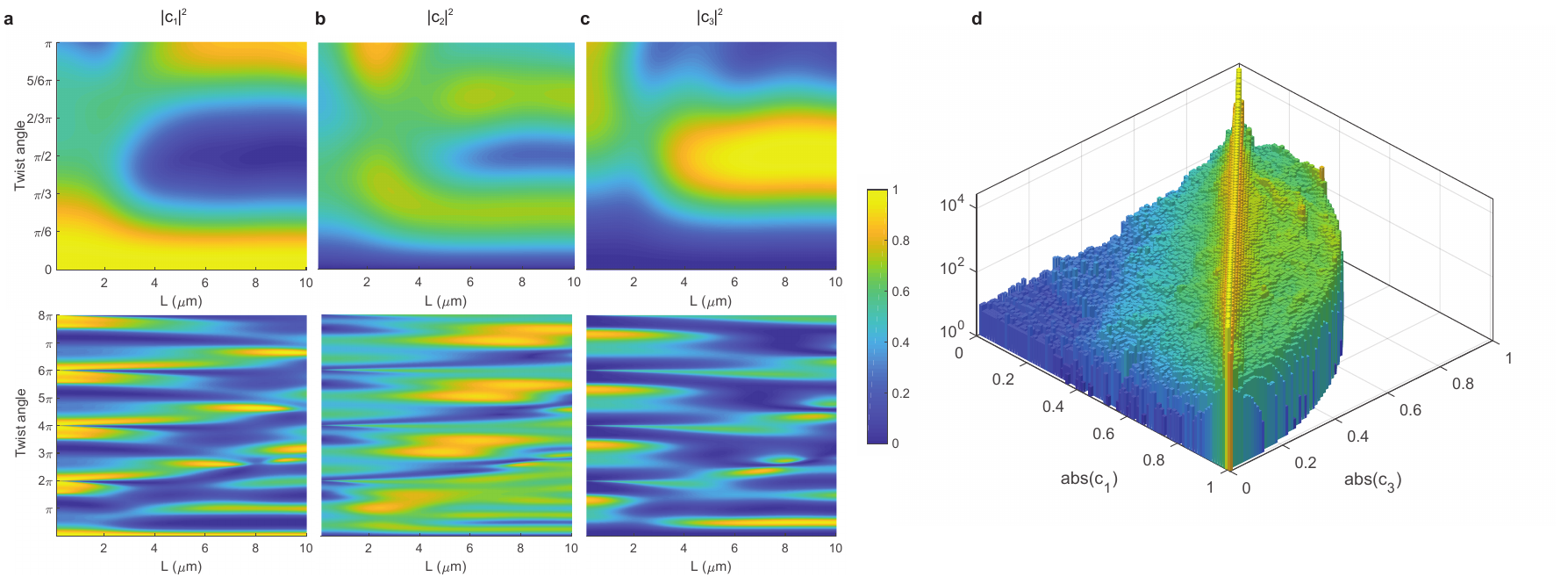}
\caption{{\bf{Calculated polarization states of photon pairs.}} The squared amplitudes represent the probabilities for each of the basis states. {\bf{(a-c)}} Values of $\vert c_1\vert^2$, $\vert c_2\vert^2$, and $\vert c_3\vert^2$ versus the sample thickness and the twist angle. Top panels present the region related to the current experiments, while the bottom panels show the effect of larger twist angles, up to 4 full twists, demonstrating the possibility to generate broader sets of polarization states. In all calculations, the pump polarization was diagonal ($D$). {\bf{d)}} Theoretical prediction on the possible states that can be produced. All possible combinations of $|c_1|$ and $|c_3|$ that can be achieved through varying the physical parameters of the sample (namely, sample length and twist angle) and pump polarization are presented. Different combinations are obtained with different occurrence rate, as certain combinations will only be satisfied by very specific set of parameters. However, this result shows that a set of parameters that can generate any desired combination always exists. Sample length up to \SI{40}{\micro m} and twist angle up to $8\pi$ is used in the simulation as well as 6 different pump polarizations, namely $H, V, A, D, R$ and $L$.}
\label{ED9_calc}
\end{figure}

\newpage
\section*{Supplementary material}
\subsection*{REFRACTIVE INDEX MEASUREMENT}
We measured the material's refractive index and birefringence with an Abbe refractometer. The material was dropcasted on the imaging slide of the refractometer without any aligning layer and left to stabilize on the refractometer for one day before the measurements. We measured the refractive index as a function of wavelength using a set of narrow bandpass filters with \SI{10}{nm} bandwidth. The ordinary value of the refractive index was estimated from the lower index boundary line, while the extraordinary value was estimated for the orthogonal incoming light polarization from the higher index boundary gradient. Additionally, the birefringence of the sample was independently measured via the wedge cell method, which showed a consistent result with the birefringence calculated from the measured refractive index values. We fitted the measured values of the refractive index with a two-term Cauchy model $n(\lambda)=A+B/\lambda^2$ to extrapolate the refractive index dispersion to the infrared region.  $A=1.663$, $B=\SI{10256}{nm^2}$ for extraordinary refractive index and $A=1.474$, $B=\SI{874}{nm^2}$ for ordinary refractive index.

\subsection*{SAMPLE CHARACTERIZATION VIA\\ SECOND-HARMONIC 
GENERATION}

We characterize the second-order nonlinearity of LC by measuring second-harmonic generation (SHG) in one of the samples and comparing it with the SHG in a known material (Extended Data Fig.~3). For comparison, we took a thin layer of 5\% magnesium-doped lithium niobate (5\% MgO:LiNbO$_3$, LN) with a thickness of \SI{7}{\micro m}. The sample under investigation was a \SI{7}{\micro m}-thick cell of FNLC-1751 with no molecular twist to avoid polarization transformation effects on SHG. Since the nonlinear tensor of LN is well known, we retrieved information about the nonlinear tensor of LC by comparing the SHG efficiency in LN and LC measured under the same experimental conditions.

As a pump, we used light generated at 1370 nm from a homemade optical parametric generator (OPG) pumped at 532 nm (20 ps pulse duration). With a set of a polarizer and a half-wave plate (HWP), installed both before and after the sample, we measured SHG in LN and LC as a function of the pump polarization and the detected second-harmonic polarization. We could retrieve the relative values of the nonlinear tensor for LC compared to LN from the obtained dependencies.

The second-order nonlinearity of any material is generally described by the $\hat{\chi}^{(2)}$ or $d$ tensor. The latter has the form~\cite{Boyd}
\begin{equation}
    \hat{d} = 
\begin{pmatrix}
d_{11} & d_{12} & d_{13} & d_{14} & d_{15} & d_{16} \\
d_{21} & d_{22} & d_{23} & d_{24} & d_{25} & d_{26} \\
d_{31} & d_{32} & d_{33} & d_{34} & d_{35} & d_{36}
\end{pmatrix}.
\end{equation}
Omitting the geometrical factors, phase-matching, and constants, the relation between the second-harmonic electric field $\vec{E}_{SH}$ and the pump electric field $\vec{E}$ is~\cite{Shoji1997}

\begin{equation}
\begin{pmatrix}
E_{\scriptscriptstyle\parallel}\\
E_o\\
E_e
\end{pmatrix}_{SHG}
=
\begin{pmatrix}
d_{11} & d_{12} & d_{13} & d_{14} & d_{15} & d_{16} \\
d_{21} & d_{22} & d_{23} & d_{24} & d_{25} & d_{26} \\
d_{31} & d_{32} & d_{33} & d_{34} & d_{35} & d_{36}
\end{pmatrix}
\begin{pmatrix}
E^2_{\scriptscriptstyle\parallel}\\
E^2_o\\
E^2_e\\
2E_oE_e\\
2E_{\scriptscriptstyle\parallel} E_e\\
2E_{\scriptscriptstyle\parallel} E_o
\end{pmatrix}_{PUMP},
\end{equation}
where indices $e$, $o$, $\parallel$ define the components of the electric field along the extraordinary axis, ordinary axis, and longitudinal component of the field, respectively. While the latter is usually equal to zero, the first two components depend on the pump polarization with respect to the orientation of the crystal axes. For LN, the d tensor has the form~\cite{Boyd}

\begin{equation}
    \hat{d}_{LN} = 
    \begin{pmatrix}
        0 & 0 & 0 & 0 & d_{31} & -d_{22} \\
        -d_{22} & d_{22} & 0 & d_{31} & 0 & 0 \\
        d_{31} & d_{31} & d_{33} & 0 & 0 & 0
    \end{pmatrix},
\end{equation}
where $d_{22} \approx 2.1~\frac{pm}{V}$,  $d_{31} \approx -4.3~\frac{pm}{V}$, and $d_{33} \approx -34~\frac{pm}{V}$~\cite{Dmitriev1999}.  If we place LN with the extraordinary axis being horizontally oriented, then the horizontal and vertical components of the generated second harmonic field $E^{(SHG)}_H$ and  $E^{(SHG)}_V$ are
\begin{gather}
    E^{(SHG)}_H = d^{(LN)}_{33}E_H^2+d^{(LN)}_{31}E_V^2,\\
    E^{(SHG)}_V = d^{(LN)}_{22}E_V^2+2d^{(LN)}_{31}E_HE_V,
\end{gather}
where $E_H = \cos\left(\theta_P\right)E_p$ and $E_V = \sin\left(\theta_P\right)E_p$ are the horizontal and vertical projections of the pump field, with $\theta_P$ being the angle between the pump polarization plane and the extraordinary axis of LN. The total detected SHG intensity is the function of the analyzer orientation given by angle $\theta_A$ between the horizontal orientation and the transmitted second-harmonic polarization,
\begin{equation}
    I_{SHG} \propto \left|E^{(SHG)}_H\cos\theta_A+E^{(SHG)}_V\sin\theta_A\right|^2.
\end{equation}
We use this equation to fit the measurement results and retrieve the values of the second-order nonlinear tensor of LC. In the experiment, the $d$ tensor of LC is defined with the extraordinary axis oriented vertically along the molecular orientation. Without any longitudinal fields involved and with no sample rotation, it is possible to retrieve 6 components of the $d$ tensor, $d_{22}$, $d_{23}$, $d_{24}$, $d_{32}$, $d_{33}$, $d_{34}$:
\begin{gather}
    E^{(SHG)}_H = d^{(LC)}_{22}E_H^2+d^{(LC)}_{23}E_V^2+2d^{(LC)}_{24}E_HE_V,\\
    E^{(SHG)}_V = d^{(LC)}_{32}E_H^2+d^{(LC)}_{33}E_V^2+2d^{(LC)}_{34}E_HE_V.
\end{gather}
although, due to uniaxial symmetry, there are complementary components with the same values. To properly compare the SHG efficiencies in LN and LC from the measured intensities, we also considered the difference in the refractive index of two materials~\cite{Shoji1997}.

First, we measure the second harmonic from LN and LC with the fixed pump polarization horizontally or vertically (panel a in Extended Data Fig.~3). For LN with the crystal axis oriented horizontally, the SHG efficiency must follow
\begin{equation}
    \eta_{SHG} = \frac{P_{SH}}{P^2_{P}} = \frac{\left(d_{33}\cos{\phi_A}\right)^2}{n^2_{LN}(\omega_{P})n_{LN}(2\omega_{P})},
\end{equation}
where indices SH, P, and A stand for second harmonic, pump, and analyzer, respectively. We used this equation to fit the measured SHG efficiency in LN from the horizontally polarized pump (red curve in Extended Data Fig.~3a) to retrieve the relative value of the $d_{33}$ component of the LN nonlinear tensor for the further comparison with the SHG efficiency in LC, which is $2.4\pm 0.03$ in arbitrary units. Further, no second harmonic was observed in LC with the vertically oriented molecules from the horizontally polarized pump. It allows us to conclude that $d_{22}$ and $d_{32}$ of the LC nonlinear tensor are close to zero. In contrast, the SHG efficiency in LC from the vertically polarized pump is comparable with the SHG efficiency in LN. From this measurement, we retrieved $d_{23}$ and $d_{33}$ of the LC nonlinear tensor by fitting the data with
\begin{equation}
    \eta_{SHG} = \frac{\left(d_{23}\cos{\phi_A}+d_{33}\sin{\phi_A}\right)^2}{n^2_{LC}(\omega_{P})n_{LC}(2\omega_{P})},
\end{equation}
with the values $d_{23} = 0.05\pm 0.02$ (assumed to be equal to zero) and $d_{33}=1.43\pm 0.02$. 

Next, we fixed the polarization of the detected second harmonic (horizontally or vertically) and changed the polarization of the pump. Again, we measured the SHG efficiency in LN as a reference (red points in Extended Data Fig.~3b). We observed no horizontally polarized second harmonic from LC (green points in Extended Data Fig.~3b), while the vertically polarized second harmonic is quite strong. From the fit of the data (blue points in Extended Data Fig.~3b) with the function
\begin{equation}
    \eta_{SHG} = \frac{\left(d_{32}\cos^2{\phi_P}+d_{33}\sin^2{\phi_P}+d_{34}\sin{2\phi_P}\right)^2}{n^2_{LC}(\omega_{P})n_{LC}(2\omega_{P})},
\end{equation}
we extracted the near-zero values for $d_{32}$ and $d_{34}$, and the similar estimation for $d_{33} = 1.46\pm 0.02$. Therefore, after the comparison of the measured values of the LC nonlinear tensor with the reference values of that for LN, we concluded that only one component of the $d$ tensor of LC is non-zero, $d_{33} \approx$ \SI{20}{\frac{pm}{V}}.

Further on, we illustrate the effect of molecular orientation switching under the applied electric field by measuring the second harmonic radiation as a function of the applied voltage (Extended Data Fig.~4, a and b). We tested the sample with the $\pi$ molecular twist pumped at \SI{850}{nm} (OPOTEK Opolette 355, \SI{5}{ns} pulse duration, \SI{20}{Hz} repetition rate). The energy of the pulses was typically in the range between \SIrange{50}{100}{\micro J}. A longpass filter with the edge at \SI{800}{nm} was used to filter out any SHG signal generated in the OPO. The pump beam was focused on the sample through a 50/50 beamsplitter and 10x, 0.3 NA objective (Nikon), which was also used to collect the reflected light consisting of both reflected laser light and generated SHG. A shortpass filter at \SI{550}{nm} rejected the reflected laser light. The collected light was analyzed by an imaging spectrometer (Andor Shamrock SR-500i) with a \SI{15}{\micro m} wide slit, a grating with 300 lines per mm, and a CCD detector with a resolution of 1600 pixels. Typical exposure times were \SIrange{0.5}{2}{s}. By observing the SH signal as a function of time and switching on the voltage, we infer that the response time is ~0.5s (Extended Data Fig.~4c). 

\subsection*{POSSIBLE EFFECTS OF THE PUMP BEAM ON THE SAMPLE AND OTHER NONLINEAR EFFECTS.}

Throughout the measurements, no influence of the excitation laser on the sample structure was observed. In the bright-field and crossed polarisation microscopy images we do not observe any changes on the sample (both momentary or permanent) caused by the pump beam. Even several hours of illumination with a few \SI{}{mW} or using the sample continuously for a couple of months did not cause any visible permanent changes. Damage threshold testing has shown that only powers above \SI{60}{mW} and illumination times of more than \SI{30}{min} caused a small permanent damage spot. Other effects, such as optical Fréedericksz transition or phase transitions due to heating, were not observed either. According to literature, these effects are typically observed in liquid crystals at several tenths of milliwatts of laser power for a diffraction-limited spot through a high NA objective \cite{transition}. In addition, the refractive index (and birefringence) of FNLCs change relatively little with the temperature \cite{FNLC_phase}. Lastly, no other nonlinear effects, such as four-wave mixing, stimulated orientational scattering, or photorefractivity, were observed, which is in accordance with previous studies where higher powers or pulsed excitations were necessary \cite{SOS, photorefractivity}.

\subsection*{MAXIMUM-LIKELIHOOD METHOD.}

Since a density matrix must be Hermitian, it can be represented as a product of two Hermitian-conjugate matrices,
\begin{equation}
    \hat{\rho} = \frac{\hat{T}^{\dagger}\left(\vec{t}\right)\hat{T}\left(\vec{t}\right)}{Tr\left(\hat{T}^{\dagger}\left(\vec{t}\right)\hat{T}\left(\vec{t}\right)\right)},
\end{equation}
where $\hat{T}\left(\vec{t}\right)$ is a semi-diagonal matrix given as a function of a real-valued vector $\vec{t}$,
\begin{equation}
    \hat{T}\left(\vec{t}\right) = 
    \begin{pmatrix}
        t_1 & t_4+\imath t_5 & t_8+\imath t_9\\
        0 & t_2 & t_6+\imath t_7\\
        0 & 0 & t_3
    \end{pmatrix},\;\vec{t}\in \mathbb{R}^{9}.
    \label{MaxLi_matrix}
\end{equation}
We can write the elements of the density matrix as a function of parameters $\vec{t}$ explicitly as
\begin{align}
\begin{cases}
    \rho_{11} &= t_1^2,\\
    \rho_{12} &= t_1t_4+\imath t_1t_5,\\
    \rho_{13} &= t_1t_8+\imath t_1t_9,\\
    \rho_{22} &= t_2^2+t_4^2+t_5^2,\\
    \rho_{23} &= t_2t_6+t_4t_8+t_5t_9+\imath\left(t_2t_7+t_4t_9-t_5t_8\right),\\
    \rho_{33} &= t_3^2+t_6^2+t_7^2+t_8^2+t_9^2,
\end{cases}
\end{align}
from which we can obtain the system of equations to find parameters $\vec{t}$,
\begin{equation}
    \begin{cases}
        t_1 = \sqrt{\rho_{11}},\\
        t_4 = \Re\left(\rho_{12}\right)/t_1,\\
        t_5 = \Im\left(\rho_{12}\right)/t_1,\\
        t_8 = \Re\left(\rho_{13}\right)/t_1,\\
        t_9 = \Im\left(\rho_{13}\right)/t_1,\\
        t_2 = \sqrt{\rho_{22} - t_4^2 - t_5^2},\\
        t_6 = \left(\Re\left(\rho_{23}\right) -t_4t_8 -t_5t_9\right)/t_2,\\
        t_7 = \left(\Im\left(\rho_{23}\right) -t_4t_9 +t_5t_8\right)/t_2,\\
        t_3 = \sqrt{\rho_{33} - t_6^2-t_7^2-t_8^2-t_9^2}.
    \end{cases}
    \label{MaxLi_init}
\end{equation}
Although system~(\ref{MaxLi_init}) has only one non-trivial solution, not all parameters $t_i$ might be purely real if the experimentally retrieved values of the density matrix $\rho^{exp}_{ij}$ are substituted into the system, meaning that the experimental density matrix is not physical. For the MaxLi method, we took the real part of the solution of system~(\ref{MaxLi_init}) as the initial guess. Then, the best fit of the density matrix is considered to be $\hat{\rho}\left(\vec{t}_{opt}\right)$ that has the form~(\ref{MaxLi_matrix}) and gives the minimum deviation from the experimentally retrieved density matrix,
\begin{gather}
    F\left(\vec{t}\right) = \sum_{i=1}^3\sum_{j\geq i}\left|\rho^{exp}_{ij} - \rho_{ij}\left(\vec{t}\right)\right|^2,\\
    F\left(\vec{t}_{opt}\right) = \min_{\vec{t}\in \mathbb{R}^{9}} F\left(\vec{t}\right).
\end{gather}


\end{document}